\documentclass[aps,pra,twocolumn,groupedaddress,showpacs]{revtex4}
\usepackage{graphicx}
\usepackage{dcolumn}
\usepackage{bm}
\begin{document}

\title{
Error free Quantum Reading by Quasi Bell State of Entangled Coherent States }

\author{Osamu Hirota}
\email{hirota@lab.tamagawa.ac.jp}
\affiliation{
 Quantum ICT Research Institute, Tamagawa University, \\
6-1-1, Tamagawa Gakuen, Machida 194-8610 Tokyo, JAPAN
}%

\date{\today}

\begin{abstract}
Nonclassical states of light field have been exploited to provide marvellous results in quantum information science. Effectiveness of nonclassical states depends on whether physical parameter as signal is continuous or digital.
Here we present an investigation on the potential of quasi Bell state of entangled 
coherent states in  quantum reading of the classical digital memory which was pioneered by Pirandola. This is a typical example of discrete quantum discrimination.
We show that the quasi Bell state gives the error free performance in the quantum reading
that cannot be obtained by any classical state.
\end{abstract}

                             
\maketitle

\section{Introduction}

In estimating an unknown parameter of quantum system, one prepares a probe light with several quantum states. To reduce the uncertainty, one can take $N$ identical measures and average the results of those measures [1]. These strategies are very attractive in quantum metrology, 
and many applications to the modern information technology have been examined [2]. 

The signals in 
almost all of the information devices are digital and the decision problem is a major subject. 
However, when one considers a quantum measurement of the classical signal of quantum systems, one needs to take into account the quantum noise effect in the measurement of the observable.
In general, limitations in the precision or error to discriminate the quantum states are imposed by such quantum noise.
The nonclassical effect may become effective in digital signal rather than continuous signal 
as in the classical information science.
Recently, a basic model of reading scheme was discussed in the context of quantum information [3,4], employing the theory of the binary quantum states discrimination  so called Helstrom bound [5].

It may be a logical consequence to discuss the above application of quantum information science.
In this context, one of key properties which make a state nonclassical is entanglement.
The quasi Bell state based on entangled coherent states has a potential of the perfect entanglement[6], and its applications to quantum teleportation and computation have been discussed[7,8,9,10].
Furthermore, the feasibility and experimental demonstration [11] of such states 
have been reported.

Thus, we are concerned with  
applications of the quasi Bell state of entangled coherent states to quantum reading where various new technologies can emerge.
In this Letter, we show the fact that one of properties of the quasi Bell state may provide the error free  quantum reading of classical digital memory.

\section{Theoretical evaluation of physical limitation and Quasi Bell states}

\subsection{Ultimate limitation in quantum measurement}

A general theory evaluating the performance of the measurement in classical and quantum  systems is 
"communication theory" including estimation and detection. 

We can divide the issue of the theoretical limit on the ultimate sensitive measurement
into signals of continuous parameter and discrete parameter. The former clarifies the precision of the measurement observable like the phase, and the latter clarifies the minimum error performance of the decision for signals in the radar or weak external force.

The quantum Cramer-Rao inequality has been formulated[5,12,13,14], in which the  bound is asymptotically achieved by the maximum likelihood estimator as well as the classical estimation theory. 
Here let $\rho (\theta)$ be the density operator of the system.
The estimation bound is given as follows:
First, the following operator equation is defined as
\begin{equation}
\frac{\partial{\rho (\theta)}}{\partial{\theta}}=
\frac{1}{2}[A\rho (\theta)+\rho (\theta)A]
\end{equation}
where $A$ is called symmetric logarithmic derivative, which is self-adjoint operator.
Then the bound is given by
\begin{equation}
(\delta\theta)^2 \ge \frac{1}{F_Q=Tr(\rho (\theta)A^2)}
\end{equation}
$F_Q$ is also called quantum Fisher information.
The above inequality defines the principally smallest possible uncertainty in estimation 
of the value of parameter.

On the other hand, in the radar detection or weak force detection, the problem is 
whether signal exists or not.
The formulation of the ultimate detection performance is called quantum
detection theory[5,12].  Especially, for the binary quantum state discrimination, the limitation can be evaluated as follows[5]:
\begin{eqnarray}
P_e &=& \xi_0 Tr\rho_0\Pi_1 + \xi_1 Tr\rho_1\Pi_0\xi_0 \nonumber \\
& &\Pi_0+\Pi_1=I,\quad \Pi_i \ge 0
\end{eqnarray}
where $\{\xi_i \}$  is a priori probability of quantum states $\{\rho_i\}$ of the system, 
$\{\Pi_i\}$ is the detection operator, respectively.
 In the case of pure states, the optimum solution is
\begin{equation}
P_e=\frac{1}{2}[1-\sqrt {1-4\xi_0\xi_1|<\psi_0|\psi_1>|^2}]
\end{equation}

\subsection{Quasi Bell state of entangled coherent states}
In general, the Bell states consist of four orthogonal states.
However, the four entangled states based on non-orthogonal states are called 
quasi Bell state.
The specific example is the following states based on coherent state:

\begin{eqnarray}
\left\{
\begin{array}{lcl}
|\Psi_1 \rangle &=& h_{1} (|\alpha \rangle_A|\alpha \rangle_B
+|-\alpha \rangle_A|-\alpha \rangle_B ) \\

|\Psi_2 \rangle &=& h_{2} (|\alpha \rangle_A|\alpha \rangle_B
-|-\alpha \rangle_A|-\alpha \rangle_B )\\

|\Psi_3 \rangle &=& h_{3} (|\alpha \rangle_A|-\alpha \rangle_B
+ |-\alpha \rangle_A|\alpha \rangle_B )\\

|\Psi_4 \rangle &=& h_{4} (|\alpha \rangle_A|-\alpha \rangle_B
-|-\alpha \rangle_A|\alpha \rangle_B)

\end{array}
\right.
\end{eqnarray}
where
$\{h_{i}\}$ are normalized constant:$h_{1}=h_{3}=1/\sqrt{2(1+\kappa^{2})}$,
$h_{2}=h_{4}=1/\sqrt{2(1-\kappa^{2})}$, and where
$ \langle \alpha | -\alpha \rangle = \kappa$ and
$\langle -\alpha | \alpha \rangle = \kappa^*$.

Some of these quasi Bell states are not orthogonal each other.
Here, if $\kappa = \kappa^*$, then the Gram matrix of them
becomes very simple as follows:
\begin{equation}
G=
\left(
\begin{array}{cccc}
1& 0& D& 0\\
0& 1& 0& 0\\
D& 0& 1& 0\\
0& 0& 0& 1\\
\end{array}
\right)
\label{ohmsgrammat}
\end{equation}
where $D=\frac{2 \kappa}{1 + {\kappa}^2}$.

  The degrees of entanglement for quasi Bell state are well known as follows:
 \begin{eqnarray}
& &E(|\Psi_1\rangle)=E(|\Psi_3\rangle) \\
                 &=& - \frac{1+C_{13}}{2} \log \frac{1+C_{13}}{2}
                   - \frac{1-C_{13}}{2} \log \frac{1-C_{13}}{2}\nonumber
\end{eqnarray}
where $C_{ij}=|\langle\Psi_i|\Psi_j\rangle|$, 
and we have the special property such as  $E(|\Psi_2\rangle)=E(|\Psi_4\rangle)=1$ [6].

Thus $|\Psi_2\rangle$ and  $|\Psi_4\rangle$ have the perfect entanglement.

\section{Specific Model of quantum reading}
There are two types of reading scheme of classical digital memory. These are  
"Amplitude shift keying (ASK)" and  "Phase shift keying(PSK)", respectively.

The problem in the current proposal of the quantum reading is to discriminate 
 quantum channels or  quantum states affected by the different reflections $r_0$ and $r_1$ to one of two modes [3,4]. This corresponds to the model of ASK scheme. 
If the reflection coefficients are  $r_0=1$ and $r_1<< 1$, it imposes great energy loss 
effect. 
The typical systems in quantum information science suffer the degradation from the energy loss effect. If one employs the nonclassical state in the above model, the loss effect has to be 
taken into account.

On the other hand, PSK scheme to realize the classical digital memory does not have a loss effect. 
We will employ PSK scheme such that the memory on the classical disk consists of the flat and concave which correspond to $"0"$ and $"1"$, respectively.
This type of memory has been invented in the beginning of the 1980's, employing a Gas laser 
with high coherence as the light source. Although, later, its scheme was replaced with ASK by employing a laser diode with low coherence, recently PSK scheme is revived by laser diode with high coherence.

The difference of two schemes is not important in the case of classical state, but they 
make great difference in the case of nonclassical states. 
 Here, desirable scheme is PSK scheme,
because one can describe the phase shift by an unitary operator as follows:
\begin{equation} 
U(\theta)=exp(-\theta {a}^{\dagger}a)
\end{equation}
where $a$ and ${a}^{\dagger}$ are the annihilation and creation operator in bosonic system, respectively.

The reading method in the current PSK scheme is to illuminate the laser light (coherent state) on a disk, and to read the phase difference. That is, the signal states of light to discriminate are
\begin{eqnarray}
|\alpha(0) \rangle &=& {\it {I}} |\alpha \rangle \\
|\alpha(1)\rangle &=& U(\theta)|\alpha \rangle
\end{eqnarray}
where ${\it {I}}$ is the identity operator.

\section{Error free quantum reading}
Let us consider the detection problem of the above model.
Here we restrict the problem to the binary detection. So detection targets are two quantum states. Since the phase difference will be $\pi$, 
the problem is to read the phase shift $\pi$ from the steady state or input state.

The coherent states are prepared in the current classical memory, and the target signal model becomes as follows:
\begin{eqnarray}
|\alpha(0) \rangle &=& {\it {I}} |\alpha \rangle=  |\alpha \rangle\\
|\alpha(1)\rangle &=& U(\theta=\pi)|\alpha \rangle = |-\alpha \rangle
\end{eqnarray}
If one employs the conventional homodyne receiver to discriminate the above quantum states,
the limitation is imposed by so called quantum shot noise limit. So this is the standard quantum limit in the modern information technology. 
If one employs the injection back effect [15] to a laser diode to detect the phase delay of the reflection from a disk as the conventional system, the sensitivity is inferior to a homodyne receiver, though it is a typical scheme.

However, in general, one can enjoy the Helstrom bound to overcome the standard quantum measurement which is achieved by a homodyne receiver. From Eq(4),  
the limitation depends on the inner product between two quantum states. For example, the inner product of two coherent states is
\begin{equation}
\langle \alpha|-\alpha\rangle =exp{(-2|\alpha|^2)}
\end{equation}

Let us assume that  $|\Psi_{2} \rangle$ of quasi Bell state in Eq(5) is employed as the light source, and 
the $B$ mode is illuminated to a memory disk. The reflection effect $U_B(\theta)$ operates on $B$ mode, so the channel model is 
\begin{equation}
{\cal {\epsilon}}_{A \otimes B} =I_A \otimes U_B(\pi)
\end{equation}

Then, the target states become as follows:
\begin{eqnarray}
|\Psi_{2}(0) \rangle &=& h_{2} (|\alpha \rangle_A|\alpha \rangle_B
-|-\alpha \rangle_A|-\alpha \rangle_B )  \nonumber \\
|\Psi_{2}(1)\rangle &=& {\cal {\epsilon}}_{A \otimes B} h_{2} (|\alpha \rangle_A|\alpha \rangle_B
-|-\alpha \rangle_A|-\alpha \rangle_B ) \nonumber \\
&=&h_{2} (|\alpha \rangle_A|-\alpha \rangle_B
-|-\alpha \rangle_A|\alpha \rangle_B ) 
\end{eqnarray} 
Thus, the input state $|\Psi_2(0)\rangle =|\Psi_2\rangle$ is 
changed to $|\Psi_2(1)\rangle = |\Psi_4\rangle$. 
The inner product between the above two entangled coherent states 
in quasi Bell state is 
\begin{equation}
\langle\Psi_{2}(0)|\Psi_{ 2}(1)\rangle= 0
\end{equation}
That is, the inner product becomes zero, and it is independent of the energy of light source.
This is a  property of the quasi Bell state.
To check this special property, one can examine the different phase shift as follows:
\begin{equation}
|\langle\Psi_{2}(0)|I_A \otimes U_B(\theta) |\Psi_{ 2}(0)\rangle |> 0
\end{equation}
where $\theta \ne \pi$.

Finally, let us employ the quantum optimum receiver for the binary pure state of two modes. The ultimate detection performances of systems with the coherent state and 
quasi Bell state of entangled coherent states are given, respectively,  as follows:

\begin{eqnarray}
& &P_e(C)=\frac{1}{2}[1-\sqrt {1-4\xi_0\xi_1 exp(-4|\alpha|^2)}]\nonumber \\
& &P_e(ECS)= 0
\end{eqnarray}
Thus, one can see that the property of the quasi Bell state provides an attractive improvement, and this property can be obtained only by the combination of the nonclassical state and the  quantum optimum receiver.
Although there are many nonclassical states, almost all is not changed 
from the input state to the orthogonal state to the original input state just by reflection.
So we interpret that such an effectiveness of the quasi Bell state comes from the special phenomena on entanglement based on non-orthogonal state, which is a feature of quasi Bell state.

\section{CONCLUSION}
In the quantum metrology, utilizing an entangled coherent state could be applicable to many super sensitive measurements. But, there is an attractive 
improvement especially in the digital quantum state discrimination rather than continuous quantum state[16].
 We have examined a potential of the entangled coherent states with 
a perfect entanglement in quasi Bell state for a well known quantum reading issue, and shown that it has the potential to realize the error free quantum reading under the finite energy of light source.

\begin{acknowledgments}
The author is grateful S.J.van Enk, Bill Munro, K.Kato, and T.Usuda for helpful discussions.
\end{acknowledgments}

\end{document}